# There are no particles, there are only fields

Art Hobson [a]
*Department of Physics, University of Arkansas, Fayetteville, AR,*
*ahobson@uark.edu*

Quantum foundations are still unsettled, with mixed effects on science and society. By now it should be possible to obtain consensus on at least one issue: Are the fundamental constituents fields or particles? As this paper shows, experiment and theory imply unbounded fields, not bounded particles, are fundamental. This is especially clear for relativistic systems, implying it's also true of non-relativistic systems. Particles are epiphenomena arising from fields. Thus the Schroedinger field is a space-filling physical field whose value at any spatial point is the probability amplitude for an interaction to occur at that point. The field for an electron *is* the electron; each electron extends over both slits in the 2-slit experiment and spreads over the entire pattern; and quantum physics is about interactions of microscopic systems with the macroscopic world rather than just about measurements. It's important to clarify this issue because textbooks still teach a particles- and measurement-oriented interpretation that contributes to bewilderment among students and pseudoscience among the public. This article reviews classical and quantum fields, the 2-slit experiment, rigorous theorems showing particles are inconsistent with relativistic quantum theory, and several phenomena showing particles are incompatible with quantum field theories.

## I. INTRODUCTION

Physicists are still unable to reach consensus on the principles or meaning of science's most fundamental and accurate theory, namely quantum physics. An embarrassment of enigmas abounds concerning wave-particle duality, measurement, nonlocality, superpositions, uncertainty, and the meaning of quantum states.[1] After over a century of quantum history, this is scandalous.[2, 3]

It's not only an academic matter. This confusion has huge real-life implications. In a world that cries out for general scientific literacy,[4] quantum-inspired pseudoscience has become dangerous to science and society. *What the Bleep Do We Know*, a popular 2004 film, won several film awards and grossed $10 million; it's central tenet is that we create our own reality through consciousness and quantum mechanics. It features physicists saying things like "The material world around us is nothing but possible movements of consciousness," it purports

to show how thoughts change the structure of ice crystals, and it interviews a 35,000 year-old spirit "channeled" by a psychic.[5] "Quantum mysticism" ostensibly provides a basis for mind-over-matter claims from ESP to alternative medicine, and provides intellectual support for the postmodern assertion that science has no claim on objective reality.[6] According to the popular television physician Deepak Chopra, "quantum healing" can cure all our ills by the application of mental power.[7] Chopra's book *Ageless Body, Timeless Mind*, a New York Times Bestseller that sold over two million copies worldwide, is subtitled *The Quantum Alternative to Growing Old*.[8] *Quantum Enigma*, a highly advertised book from Oxford University Press that's used as a textbook in liberal arts physics courses at the University of California and elsewhere, bears the sub-title *Physics Encounters Consciousness*.[9] It's indeed scandalous when librarians and book store managers wonder whether to shelve a book under "quantum physics," "religion," or "new age." For further documentation of this point, see the Wikipedia article "Quantum mysticism" and references therein.

Here, I'll discuss just one fundamental quantum issue: field-particle (or wave-particle) duality. More precisely, this paper answers the following question: Based on standard non-relativistic and relativistic quantum physics, do experiment and theory lead us to conclude that the universe is ultimately made of fields, or particles, or both, or neither? There are other embarrassing quantum enigmas, especially the measurement problem, as well as the ultimate ontology (i.e. reality) implied by quantum physics. This paper studies only field-particle duality. In particular, it's neutral on the interpretations (e.g. many worlds) and modifications (e.g. hidden variables, objective collapse theories) designed to resolve the measurement problem.

Many textbooks and physicists apparently don't realize that a strong case, supported by leading quantum field theorists,[10, 11 12, 13, 14, 15, 16, 17] for a pure fields view has developed during the past three decades. Three popular books are arguments for an all-fields perspective.[18, 19, 20] I've argued the advantages of teaching non-relativistic quantum physics (NRQP, or "quantum mechanics") from an all-fields perspective;[21] my conceptual physics textbook for non-science college students assumes this viewpoint.[22]

There is plenty of evidence today for physicists to come to a consensus supporting an all-fields view. Such a consensus would make it easier to resolve other quantum issues. But fields-versus-particles is still alive and kicking, as you can see by noting that "quantum field theory" (QFT) and "particle physics" are interchangeable names for the same discipline! And there's a huge gap between the views of leading quantum physicists (Refs. 10-18) and virtually every quantum physics textbook.

Physicists are schizophrenic about fields and particles. At the high-energy end, most quantum field theorists agree for good reasons (Secs. III, V, VI) that relativistic quantum physics is about fields and that electrons, photons, and so forth are merely excitations (waves) in the fundamental fields. But at the low-energy end, most NRQP education and popular talk is about particles. Working physicists, teachers, and NRQP textbooks treat electrons, photons, protons, atoms, etc. as particles that exhibit paradoxical behavior. Yet NRQP is the non-relativistic limit of the broader relativistic theory, namely QFTs that for all the world appear to be about fields. If QFT is about fields, how can its restriction to non-relativistic phenomena be about particles? Do infinitely extended fields turn into bounded particles as the energy drops?

As an example of the field/particle confusion, the 2-slit experiment is often considered paradoxical, and it is a paradox if one assumes that the universe is made of particles. For Richard Feynman, this paradox was unavoidable. Feynman was a particles guy. As Frank Wilczek puts it, "uniquely (so far as I know) among physicists of high stature, Feynman hoped to remove field-particle dualism by getting rid of the fields " (Ref. 16). As a preface to his lecture about this experiment, Feynman advised his students,

> Do not take the lecture too seriously, feeling that you really have to understand in terms of some model what I am going to describe, but just relax and enjoy it. I am going to tell you what nature behaves like. If you will simply admit that maybe she does behave like this, you will find her a delightful, entrancing thing. Do not keep saying to yourself, if you can possibly avoid it, "But how can it be like that?" because you will get "down the drain," into a blind alley from which nobody has yet escaped. Nobody knows how it can be like that.[23]

There are many interpretational difficulties with the 2-slit experiment, and I'm certainly not going to solve all of them here. But the puzzle of wave-particle duality in this experiment can be resolved by switching to an all-fields perspective (Sect. IV).

Physics education is affected directly, and scientific literacy indirectly, by what textbooks say about wave-particle duality and related topics. To find out what textbooks say, I perused the 36 textbooks in my university's library having the words "quantum mechanics" in their title and published after 1989. 30 implied a universe made of particles that sometimes act like fields, 6 implied the fundamental constituents behaved sometimes like particles and sometimes like fields, and none viewed the universe as made of fields that sometimes appear to be particles. Yet the leading quantum field theorists argue explicitly for the latter view (Refs. 10-18). Something's amiss here.

The purpose of this paper is to assemble the strands of the fields-versus-particles discussion in order to hasten a consensus that will resolve the wave-particle paradoxes while bringing the conceptual structure of quantum physics into agreement with the requirements of special relativity and the views of leading quantum field theorists. Section II argues that Faraday, Maxwell, and Einstein viewed classical electromagnetism as a field phenomenon. Section III argues that quantum field theory developed from classical electrodynamics and then extended the quantized field notion to matter. Quantization introduced certain particle-like characteristics, namely individual quanta that could be counted, but the theory describes these quanta as extended disturbances in space-filling fields. Section IV analyzes the 2-slit experiment to illustrate the necessity for an all-fields view of NRQP. The phenomena and the theory lead to paradoxes if interpreted in terms of particles, but are comprehensible in terms of fields. Section V presents a rigorous theorem due to Hegerfeldt showing that, even if we assume a very broad definition of "particle" (namely that a particle should extend over only a finite, not infinite, region), particles contradict both relativity and quantum physics. Section VI argues that quantized fields imply a quantum vacuum that contradicts an all-particles view while confirming the field view. Furthermore, two vacuum effects--the Unruh effect and single-quantum nonlocality--imply a field view. Thus, many lines of reasoning contradict the particles view and confirm the field view. Section VII summarizes the conclusions.

## II. A HISTORY OF CLASSICAL FIELDS

Fields are one of physics' most plausible notions, arguably more intuitively credible than pointlike particles drifting in empty space. It's perhaps surprising that, despite the complete absence of fields from Isaac Newton's *Principia* (1687), Newton's intuition told him the universe is filled with fields. In an exchange of letters with Reverend Richard Bentley explaining the *Principia* in non-scientists' language, Newton wrote:

> It is inconceivable that inanimate brute matter should, without the mediation of something else which is not material, operate upon and affect other matter without mutual contact… That gravity should be innate, inherent, and essential to matter, so that one body may act upon another at a distance through a vacuum, without the mediation of anything else, by and through which their action and force may be conveyed from one to another, is to me so great an absurdity that I believe no man who has in philosophical matters a competent faculty of thinking can ever fall into it.[24]

But Newton couldn't find empirical evidence to support a causal explanation of gravity, and any explanation remained purely hypothetical. When writing or speaking of a possible underlying mechanism for gravity, he chose to remain silent, firmly maintaining "I do not feign hypotheses" (Ref. 18, p. 138). Thus it was generally accepted by the beginning of the 19th century that a fundamental physical theory would contain equations for direct forces-at-a-distance between tiny indestructible atoms moving through empty space. Before long, however, electromagnetism and relativity would shift the emphasis from action-at-a-distance to fields.

The shift was largely due to Michael Faraday (1791-1867). Working about 160 years after Newton, he introduced the modern concept of fields as properties of space having physical effects.[25] Faraday argued against action-at-a-distance, proposing instead that interactions occur via space-filling "lines of force" and that atoms are mere convergences of these lines of force. He recognized that a demonstration that non-instantaneous electromagnetic (EM) interactions would be fatal to action-at-a-distance because interactions would then proceed gradually from one body to the next, suggesting that some physical process occurred in the intervening space. He saw lines of force as space-filling physical entities that could move, expand, and contract. He concluded that magnetic lines of force, in particular, are physical conditions of "mere space" (i.e. space containing no material substance). Today this description of fields as "conditions of space" is standard.[26]

James Clerk Maxwell (1831-1879) was less visionary, more Newtonian, and more mathematical than Faraday. By invoking a mechanical ether that obeyed Newton's laws, he brought Faraday's conception of continuous transmission of forces rather than instantaneous action-at-a-distance into the philosophical framework of Newtonian mechanics. Thus Faraday's lines of force became the state of a material medium, "the ether," much as a velocity field is a state of a material fluid. He found the correct dynamical field equations for EM phenomena, consistent with all known experimental results. His analysis led to the predictions of (1) a finite transmission time for EM actions, and (2) light as an EM field phenomenon. Both were later spectacularly confirmed. Despite the success of his equations, and despite the non-appearance of ether in the actual equations, Maxwell insisted throughout his life that Newtonian mechanical forces in the ether produce all electric and magnetic phenomena, a view that differed crucially from Faraday's view of the EM field as a state of "mere space."

Experimental confirmations of the field nature of light, and of a time delay for EM actions, were strong confirmations of the field view. After all, light certainly seems real. And a time delay demands the presence of energy in the intervening space in order to conserve energy. That is, if energy is emitted here and now, and received there and later, then where is it in the meantime? Clearly, it's in the field.[27]

Faraday and Maxwell created one of history's most telling changes in our physical worldview: the change from particles to fields. As Albert Einstein put it, "Before Maxwell, Physical Reality …was thought of as consisting in material particles…. Since Maxwell's time, Physical Reality has been thought of as represented by continuous fields, ...and not capable of any mechanical interpretation. This change in the conception of Reality is the most profound and the most fruitful that physics has experienced since the time of Newton."[28]

As the preceding quotation shows, Einstein supported a "fields are all there is" view of classical (but not necessarily quantum) physics. He put the final logical touch on classical fields in his 1905 paper proposing the special theory of relativity, where he wrote "The introduction of a 'luminiferous' ether will prove to be superfluous."[29] For Einstein, there was no material ether to support light waves. Instead, the "medium" for light was space itself. That is, for Einstein, fields are states or conditions of *space*. This is the modern view. The implication of special relativity (SR) that energy has inertia further reinforces both Einstein's rejection of the ether and the significance of fields. Since fields have energy, they have inertia and should be considered "substance like" themselves rather than simply states of some substance such as ether.

The general theory of relativity (1916) resolves Newton's dilemma concerning the "absurdity" of gravitational action-at-a-distance. According to general relativity, the universe is full of gravitational fields, and physical processes associated with this field occur even in space that is free from matter and EM fields. Einstein's field equations of general relativity are

$$R_{\mu\nu}(x) - (1/2)g_{\mu\nu}(x)R(x) = T_{\mu\nu}(x) , \qquad (1)$$

where $x$ represents space-time points, $\mu$ and $\nu$ run over the 4 dimensions, $g_{\mu\nu}(x)$ is the metric tensor field, $R_{\mu\nu}(x)$ and $R(x)$ are defined in terms of $g_{\mu\nu}(x)$, and $T_{\mu\nu}(x)$ is the energy-momentum tensor of matter. These field (because they hold at every $x$) equations relate the geometry of space-time (left-hand side) to the energy and momentum of matter (right-hand side). The gravitational field is described solely by the metric tensor $g_{\mu\nu}(x)$. Einstein referred to the left-hand side of Eq. (1) as "a palace of gold" because it represents a condition of space-time and to the right-hand side as "a hovel of wood" because it represents a condition of matter.

Thus by 1915 classical physics described all known forces in terms of fields--conditions of space--and Einstein expressed dissatisfaction that matter couldn't be described in the same way.

## III. A HISTORY AND DESCRIPTION OF QUANTUM FIELDS

From the early Greek and Roman atomists to Newton to scientists such as Dalton, Robert Brown, and Rutherford, the microscopic view of matter was always dominated by particles. Thus the non-relativistic quantum physics of matter that developed in the mid-1920s was couched in particle language, and quantum physics was called "quantum mechanics" in analogy with the Newtonian mechanics of indestructible particles in empty space.[30] But ironically, the central equation of the quantum physics of matter, the Schroedinger equation, is a field equation. Rather than an obvious recipe for particle motion, it appears to describe a time-dependent field $\Psi(\mathbf{x},t)$ throughout a spatial region. Nevertheless, this field picked up a particle interpretation when Max Born proposed that $\Psi(\mathbf{x}_0,t)$ is the probability amplitude that, upon measurement at time $t$, the presumed particle "will be found" at the point $\mathbf{x}_0$. Another suggestion, still in accord with the Copenhagen interpretation but less confining, would be that $\Psi(\mathbf{x}_0,t)$ is the probability amplitude for an interaction to occur at $\mathbf{x}_0$. This preserves the Born rule while allowing either a field or particle interpretation.

In the late 1920s, physicists sought a relativistic theory that incorporated quantum principles. EM fields were not described by the nonrelativistic Schroedinger equation; EM fields spread at the speed of light, so any quantum theory of them must be relativistic. Such a theory must also describe emission (creation) and absorption (destruction) of radiation. Furthermore, NRQP says energy spontaneously fluctuates, and SR ($E=mc^2$) says matter can be created from non-material forms of energy, so a relativistic quantum theory must describe creation and destruction of matter. Schroedinger's equation needed to be generalized to include such phenomena. QFTs, described in the remainder of this Section, arose from these efforts.

### A. Quantized radiation fields

"How can any physicist look at radio or microwave antennas and believe they were meant to capture particles?"[31] It's implausible that EM signals transmit from antenna to antenna by emitting and absorbing particles; how do antennas "launch" or "catch" particles? In fact, how do signals transmit? Instantaneous transmission is ruled out by the evidence. Delayed transmission by direct action-at-a-distance without an intervening medium has been tried in theory and found wanting.[32] The 19-century answer was that transmission occurs via the EM field. Quantum physics preserves this notion, while "quantizing" the field. The field itself remains continuous, filling all space.

The first task in developing a relativistic quantum theory was to describe EM radiation--an inherently relativistic phenomenon--in a quantum fashion. So it's not surprising that QFT began with a quantum theory of radiation.[33 34 35] This problem was greatly simplified by the Lorentz covariance of Maxwell's equations--they satisfy SR by taking the same form in every inertial reference frame. Maxwell was lucky, or brilliant, in this regard.

A straightforward approach to the quantization of the "free" (no source charges or currents) EM field begins with the classical vector potential field $\boldsymbol{A}(\boldsymbol{x},t)$ from which we can derive $\boldsymbol{E}(\boldsymbol{x},t)$ and $\boldsymbol{B}(\boldsymbol{x},t)$.[36] Expanding this field in the set of spatial fields $exp(\pm i\boldsymbol{k}\bullet\boldsymbol{x})$ (orthonormalized in the delta-function sense for large spatial volumes) for each vector $\boldsymbol{k}$ having positive components,

$$\boldsymbol{A}(\boldsymbol{x},t) = \sum_k [\boldsymbol{a}(\boldsymbol{k},t)\, exp(i\boldsymbol{k}\bullet\boldsymbol{x}) + \boldsymbol{a}^*(\boldsymbol{k},t)\, exp(-i\boldsymbol{k}\bullet\boldsymbol{x})]. \qquad (2)$$

The field equation for $\boldsymbol{A}(\boldsymbol{x},t)$ then implies that each coefficient $\boldsymbol{a}(\boldsymbol{k},t)$ satisfies a classical harmonic oscillator equation. One regards these equations as the equations of motion for a mechanical system having an infinite number of degrees of freedom, and quantizes this classical mechanical system by assuming the $\boldsymbol{a}(\boldsymbol{k},t)$ are operators $\boldsymbol{a}_{op}(\boldsymbol{k},t)$ satisfying appropriate commutation relations and the $\boldsymbol{a}^*(\boldsymbol{k},t)$ are their adjoints. The result is that Eq. (2) becomes a vector operator-valued field

$$\boldsymbol{A}_{op}(\boldsymbol{x},t) = \sum_k [\boldsymbol{a}_{op}(\boldsymbol{k},t)\, exp(i\boldsymbol{k}\bullet\boldsymbol{x}) + \boldsymbol{a}^*_{op}(\boldsymbol{k},t)\, exp(-i\boldsymbol{k}\bullet\boldsymbol{x})], \qquad (3)$$

in which the amplitudes $\boldsymbol{a}^*_{op}(\boldsymbol{k},t)$ and $\boldsymbol{a}_{op}(\boldsymbol{k},t)$ of the **k**th "field mode" satisfy the Heisenberg equations of motion (in which the time-dependence resides in the operators while the system's quantum state $|\Psi>$ remains fixed) for a set of quantum harmonic oscillators. Assuming $\boldsymbol{a}^*_{op}$ and $\boldsymbol{a}_{op}$, commute for bosons, one can show that these operators are the familiar raising and lowering operators from the harmonic oscillator problem in NRQP. $\boldsymbol{A}_{op}(\boldsymbol{x},t)$ is now an operator-valued field whose dynamics obey quantum physics. Since the classical field obeyed SR, the quantized field satisfies quantum physics and SR.

Thus, as in the harmonic oscillator problem, the $\boldsymbol{k}$th mode has an infinite discrete energy spectrum $hf_{\boldsymbol{k}}(N_{\boldsymbol{k}}+1/2)$ with $N_{\boldsymbol{k}} = 0, 1, 2, ...$ where $f_{\boldsymbol{k}} = c|\boldsymbol{k}|/2\pi$ is the mode's frequency.[37] As Max Planck had hypothesized, the energy of a single mode has an infinite spectrum of discrete possible values separated by $\Delta E=hf_{\boldsymbol{k}}$. $N_{\boldsymbol{k}}$ is the number of Planck's energy bundles or quanta in the **k**th mode. Each quantum is called an "excitation" of the field, because its energy $hf_k$ represents additional field energy. EM field quanta are called "photons," from the Greek word for light. A distinctly quantum aspect is that, even in the vacuum state where $N_k = 0$, each mode has energy $hf_k/2$. This is because the $\boldsymbol{a(k, t)}$ act like quantum harmonic oscillators, and these must have energy even in the ground state because of the uncertainty principle. Another quantum aspect is that EM radiation is "digitized" into discrete quanta of energy $hf$. You can't have a fraction of a quantum.

Because it defines an operator for every point $\boldsymbol{x}$ throughout space, the operator-valued field Eq. (3) is properly called a "field." Note that, unlike the NRQP case, $\boldsymbol{x}$ is not an operator but rather a parameter, putting $\boldsymbol{x}$ on an equal footing with $t$ as befits a relativistic theory. E.g. we can speak of the expectation value of the field $\boldsymbol{A}_{op}$ at $\boldsymbol{x}$ and $t$, but we cannot speak of the expectation value of $\boldsymbol{x}$ because $\boldsymbol{x}$ is not an observable. This is because fields are inherently extended in space and don't have specific positions.

But what does the operator field Eq. (3) operate on? Just as in NRQP, operators operate on the system's quantum state $|\Psi\rangle$. But the Hilbert space for such states cannot have the same structure as for the single-body Schroedinger equation, or even its N-body analog, because N must be allowed to vary in order to describe creation and destruction of quanta. So the radiation field's quantum states exist in a Hilbert space of variable N called "Fock space." Fock space is the (direct) sum of N-body Hilbert spaces for N = 0, 1, 2, 3, .... Each component N-body Hilbert space is the properly symmetrized (for bosons or fermions) product of N single-body Hilbert spaces. Each normalized component has its own complex amplitude, and the full state $|\Psi\rangle$ is (in general) a superposition of states having different numbers of quanta.

An important feature of QFT is the existence of a vacuum state $|0\rangle$, a unit vector that must not be confused with the zero vector (having "length" zero in Fock space), having no quanta ($N_k=0$ for all $\boldsymbol{k}$). Each mode's vacuum state has energy $hf_k/2$. The vacuum state manifests itself experimentally in many ways, which would be curious if particles were really fundamental because there are no particles (quanta) in this state. We'll expand on this particular argument in Sect. VI.

The operator field Eq. (3) (and other observables such as energy) operates on $|\Psi\rangle$, creating and destroying photons. For example, the expected value of the vector potential is a vector-valued relativistic field $\boldsymbol{A}(\boldsymbol{x},t) = \langle \boldsymbol{A}_{op}(\boldsymbol{x},t)\rangle = \langle\Psi| \boldsymbol{A}_{op}(\boldsymbol{x},t) |\Psi\rangle$, an expression in which $\boldsymbol{A}_{op}(\boldsymbol{x},t)$ operates on $|\Psi\rangle$. We see again that $\boldsymbol{A}_{op}(\boldsymbol{x},t)$ is actually a physically meaningful field because it has a physically measurable expectation value at every point $\boldsymbol{x}$ throughout a region of space. So a classical field that is quantized does not cease to be a field.

Some authors conclude, incorrectly, that the countability of quanta implies a particle interpretation of the quantized system.[38] Discreteness is a necessary *but not sufficient* condition for particles. Quanta are countable, but they are spatially extended and certainly not particles. Eq. (3) implies that a single mode's spatial dependence is sinusoidal and fills all space, so that adding a monochromatic quantum to a field uniformly increases the *entire* field's energy (uniformly distributed throughout all space!) by *hf*. This is nothing like adding a particle. Quanta that are superpositions of different frequencies can be more spatially bunched and in this sense more localized, but they are always of infinite extent. So it's hard to see how photons could be particles.

Phenomena such as "particle" tracks in bubble chambers, and the small spot appearing on a viewing screen when a single quantum interacts with the screen, are often cited as evidence that quanta are particles, but these are insufficient evidence of particles [39, 40] (Sec. IV). In the case of radiation, it's especially difficult to argue that the small interaction points are evidence that a particle impacted at that position because photons never have positions--position is not an observable and photons cannot be said to be "at" or "found at" any particular point.[41, 42, 43, 44, 45] Instead, the spatially extended radiation field interacts with the screen in the vicinity of the spot, transferring one quantum of energy to the screen.

## B. Quantized matter fields.

QFT puts matter on the same all-fields footing as radiation. This is a big step toward unification. In fact, it's a general principal of all QFTs that fields are all there is (Refs. 10-21). For example the Standard Model, perhaps the most successful scientific theory of all time, is a QFT. But if fields are all there is, where do electrons and atoms come from? QFT's answer is that they are field quanta, but quanta of matter fields rather than quanta of force fields.[46]

"Fields are all there is" suggests beginning the quantum theory of matter from Schroedinger's equation, which mathematically is a field equation similar to Maxwell's field equations, and quantizing it. But you can't create a relativistic theory (the main purpose of QFT) this way because Schroedinger's equation is not Lorentz covariant. Dirac invented, for just this purpose, a covariant generalization of Schroedinger's equation for the field $\Psi(\mathbf{x},t)$ associated with a single electron.[47] It incorporates the electron's spin, accounts for the electron's magnetic moment, and is more accurate than Schroedinger's equation in predicting the hydrogen atom's spectrum. It however has undesirable features such as the existence of non-physical negative-energy states. These can be overcome by treating Dirac's equation as a classical field equation for matter analogous to Maxwell's equations for radiation, and quantizing it in the manner outlined in Sec. III A. The resulting quantized matter field $\Psi_{op}(\mathbf{x},t)$ is called the "electron-positron field." It's an operator-valued field operating in the anti-symmetric Fock space. Thus the non-quantized Dirac equation describes a matter field occupying an analogous role in the QFT of matter to the role of Maxwell's equations in the QFT of radiation (Refs. 12, 45). The quantized theory of electrons comes out looking similar to the preceding QFT of the EM field, but with material quanta and with field operators that now create or destroy these quanta in quantum-antiquantum pairs (Ref. 36).

It's not difficult to show that standard NRQP is a special case, for non-relativistic material quanta, of relativistic QFT (Ref. 36). Thus the Schroedinger field is the non-relativistic version of the Dirac equation's relativistic field. It follows that the Schroedinger matter field, the analogue of the classical EM field, is a physical, space-filling field. Just like the Dirac field, this field *is* the electron.

## C. Further properties of quantum fields

Thus the quantum theory of electromagnetic radiation is a re-formulation of classical electromagnetic theory to account for quantization--the "bundling" of radiation into discrete quanta. It remains, like the classical theory, a field theory. The quantum theory of matter introduces the electron-positron field and a new field equation, the Dirac equation, the analog for matter of the classical Maxwell field equations for radiation. Quantization of the Dirac equation is analogous to quantization of Maxwell's equations, and the result is the quantized electron-positron field. The Schroedinger equation, the non-relativistic version of the Dirac equation, is thus a field equation. There are no particles in any of this, there are only field quanta--excitations in spatially extended continuous fields.

For over three decades, the Standard Model--a QFT--has been our best theory of the microscopic world. It's clear from the structure of QFTs (Secs. III A and III B) that they actually are field theories, not particle theories in disguise. Nevertheless, I'll offer further evidence for their field nature here and in Secs. V and VI.

Quantum fields have one particle-like property that classical fields don't have: They are made of countable quanta. Thus quanta cannot partly vanish but must (like particles) be entirely and instantly created or destroyed. Quanta carry energy and momenta and can thus "hit like a particle." Following three centuries of particle-oriented Newtonian physics, it's no wonder that it took most of the 20th century to come to grips with the field nature of quantum physics.

Were it not for Newtonian preconceptions, quantum physics might have been recognized as a field theory by 1926 (Schroedinger's equation) or 1927 (QFT). The superposition principle should have been a dead giveaway: A sum of quantum states is a quantum state. Such superposition is characteristic of all linear wave theories and at odds with the generally non-linear nature of Newtonian particle physics.

A benefit of QFTs is that quanta of a given field must be identical because they are all excitations of the same field, somewhat as two ripples on the same pond are in many ways identical. Because a single field explains the existence and nature of gazillions of quanta, QFTs represent an enormous unification. The universal electron-positron field, for example, explains the existence and nature of all electrons and all positrons.

When a field changes its energy by a single quantum, it must do so instantaneously, because a non-instantaneous change would imply that, partway through the change, the field had gained or lost only a fraction of a quantum. Such fractions are not allowed because energy is quantized. Field quanta have an all-or-nothing quality. The QFT language of creation and annihilation of quanta expresses this nicely. A quantum is a unified entity even though its energy might be spread out over light years--a feature that raises issues of nonlocality intrinsic to the quantum puzzle.

"Fields are all there is" should be understood literally. For example, it's a common misconception to imagine a tiny particle imbedded somewhere in the Schroedinger field. There is no particle. An electron *is* its field.

As is well known, Einstein never fully accepted quantum physics, and spent the last few decades of his life trying to explain all phenomena, including quantum phenomena, in terms of a *classical* field theory. Nevertheless, and although Einstein would not have agreed, it seems to me that QFT achieves Einstein's dream to regard nature as fields. QFT promotes the right-hand side of Eq. (1) to field status. But it is not yet a "palace of gold" because Einstein's goal of explaining all fields entirely in terms of zero-rest-mass fields such as the gravitational field has not yet been achieved, although the QFT of the strong force comes close to this goal of "mass without mass" (Refs. 13, 16, 17).

## IV. THE 2-SLIT EXPERIMENT

### A. Phenomena

Field-particle duality appears most clearly in the context of the time-honored 2-slit experiment, which Feynman claimed "contains the *only* mystery." [48, 49] Figures 1 and 2 show the outcome of the 2-slit experiment using a dim light beam (Fig. 1) and a "dim" electron beam (Fig. 2) as sources, with time-lapse photography. The set-up is a source emitting monochromatic light (Fig. 1) or mono-energetic electrons (Fig. 2), an opaque screen with two parallel slits, and a detection screen with which the beam collides. In both figures, particle-like impacts build up on the detection screen to form interference patterns. The figures show both field aspects (the extended patterns) and particle aspects (the localized impacts). The similarity between the two figures is striking and indicates a fundamental similarity between photons and electrons. It's intuitively hard to believe that one figure was made by waves and the other by particles.

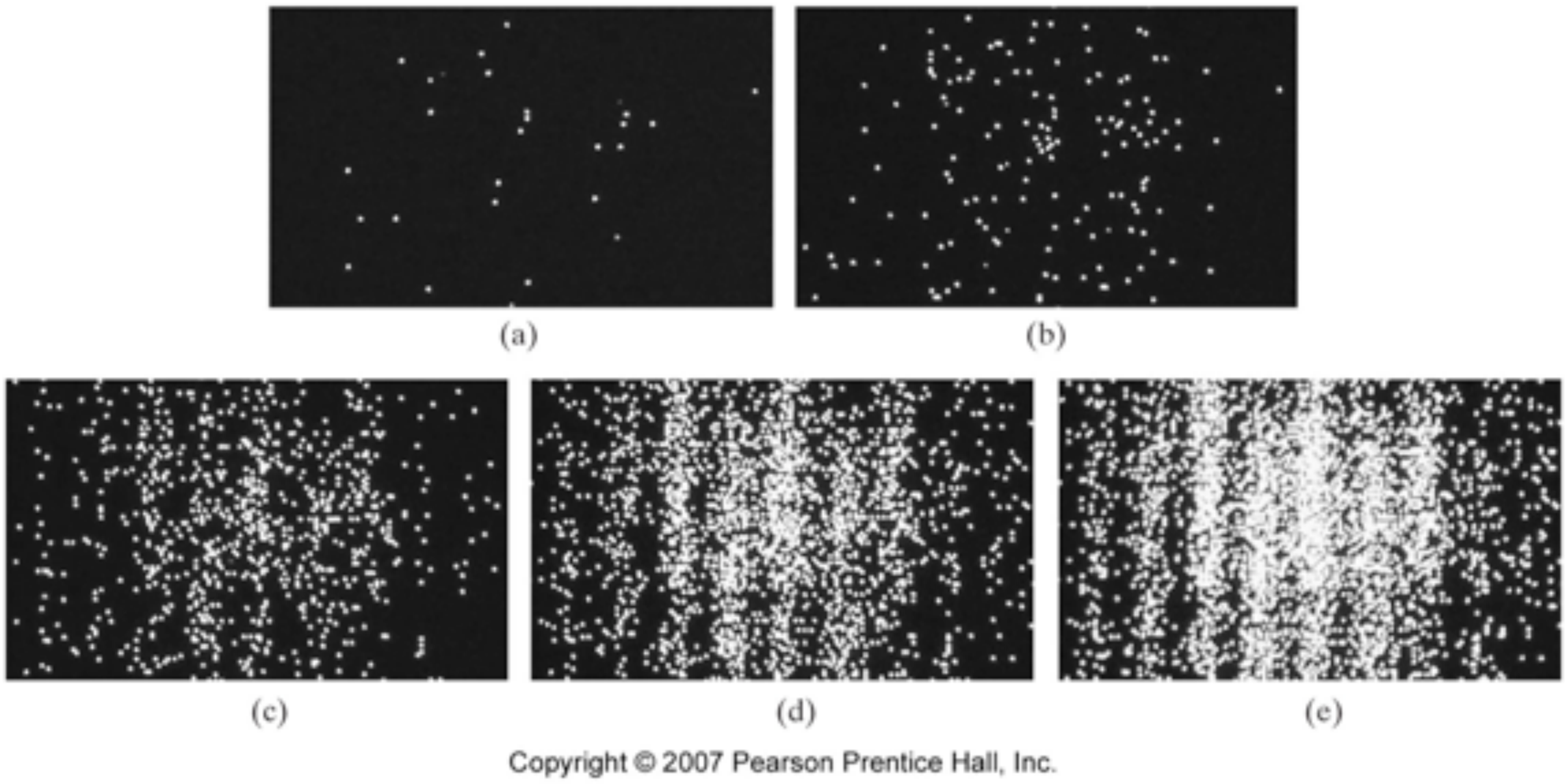

Fig. 1: The 2-slit experiment outcome using dim light with time-lapse photography. In successive images, an interference pattern builds up from particle-like impacts.[50]

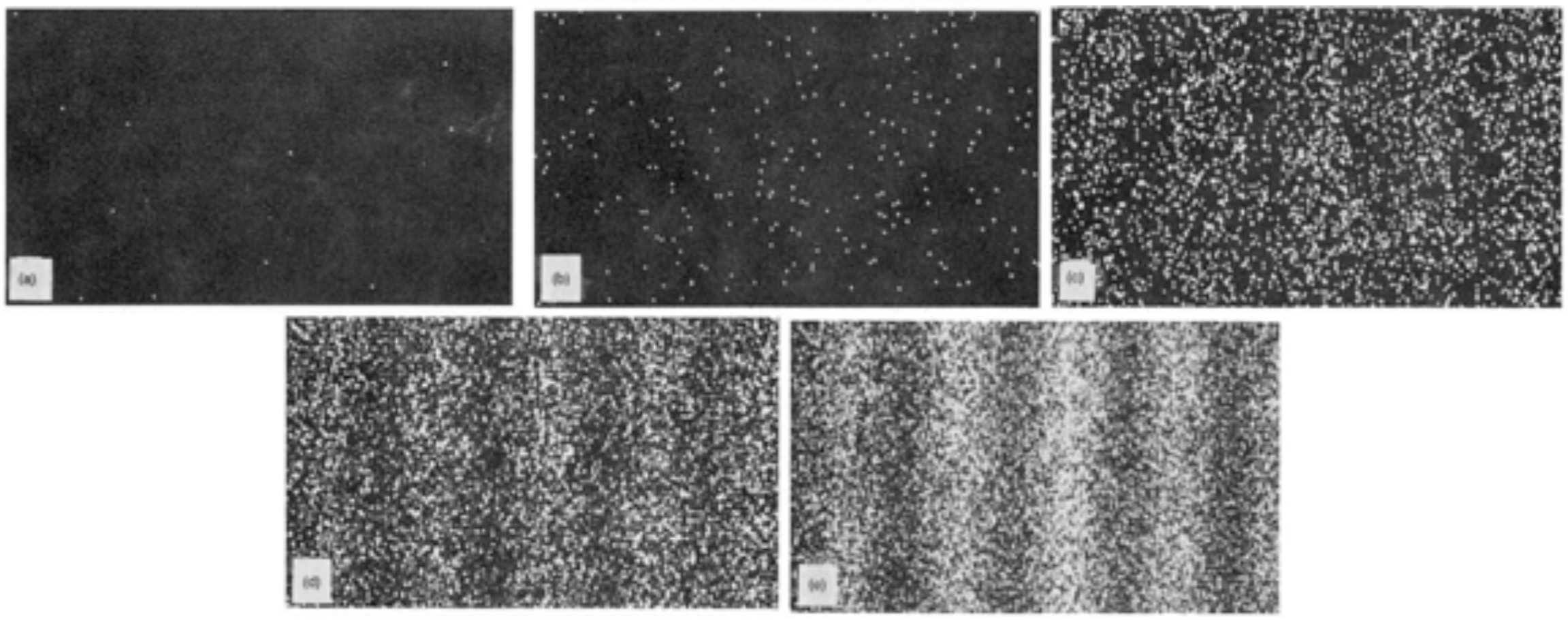

Fig 2: The 2-slit experiment outcome using a "dim" electron beam with time-lapse photography. An interference pattern builds up from particle-like impacts.[51]

Consider, first, the extended pattern. It's easy to explain if each quantum (photon or electron) is an extended field that comes through both slits. But could the pattern arise from particles? The experiments can be performed using an ensemble of separately emitted individual quanta, implying the results cannot arise from interactions between different quanta.[52] Preparation is identical for all the quanta in the ensemble. Thus, given this particular experimental context (namely the 2-slit experiment with both slits open, no detector at the slits, and a "downstream" screen that detects interactions of each ensemble member), each quantum must carry information about the entire pattern that appears on the screen (in order, e.g., to avoid all the nodes). In this sense, each quantum can be said to be spread out over the pattern.

If we close one slit, the pattern shifts to the single-slit pattern behind the open slit, showing no interference. Thus each quantum carries different information depending on whether two or one slits are open.

How does one quantum get information as to how many slits are open? If a quantum is a field that is extended over both slits, there's no problem. But could a particle coming through just one slit obtain this information by detecting physical forces from the other, relatively distant, slit? The effect is the same for photons and electrons, and the experiment has been done with neutrons, atoms, and many molecular types, making it difficult to imagine gravitational, EM, or nuclear forces causing such a long-distance force effect. What more direct evidence could there be that a quantum is an extended field? Thus we cannot explain the extended patterns by assuming each quantum is a particle, but we can explain the patterns by assuming each quantum is a field. [53]

Now consider the particle-like small impact points. We can obviously explain these if quanta are particles, but can we explain them with fields? The flashes seen in both figures are multi-atom events initiated by interactions of a single quantum with the screen. In Fig. 2, for example, each electron interacts with a portion of a fluorescent film, creating some 500 photons; these photons excite a photo cathode, producing photo electrons that are then focused into a point image that is displayed on a TV monitor (Ref. 51). This shows that a quantum can interact locally with atoms, but it doesn't show that quanta are point particles. A large object (a big balloon, say) can interact quite locally with another object (a tiny needle, say). The localization seen in the two figures is characteristic of the *detector*, which is made of localized atoms, rather than of the detected quanta. The detection, however, localizes ("collapses"--Secs. IV B and IV C) the quantum.

Similar arguments apply to the observation of thin particle tracks in bubble chambers and other apparent particle detections. Localization is characteristic of the detection process, not of the quantum that is being detected.

Thus the interference patterns in Figs. 1 and 2 confirm field behavior and rule out particle behavior, while the small interaction points neither confirm particle behavior nor rule out field behavior. The experiment thus confirms field behavior. As Dirac famously put it in connection with experiments of the 2-slit type, "The new theory [namely quantum mechanics], which connects the wave function with probabilities for one photon, gets over the difficulty [of explaining the interference] by making each photon go partly into each of the two components. Each photon then interferes only with itself." [54] The phrases in square brackets are mine, not Dirac's.

Given the extended field nature of each electron, Fig. 2 also confirms von Neumann's famous collapse postulate:[55] Each electron carries information about the entire pattern and collapses to a much smaller region upon interaction. Most textbooks set up a paradox by explicitly or implicitly assuming each quantum to come through one or the other slit, and then struggle to resolve the paradox. But if each quantum comes through both slits, there's no paradox.

## B. Theory, at the slits [56, 57]

Now assume detectors are at each slit so that a quantum passing through slit 1 (with slit 2 closed) triggers detector 1, and similarly for slit 2. Let $|\psi_1>$ and $|\psi_2>$, which we assume form an orthonormal basis for the quantum's Hilbert space, denote the states of a quantum passing through slit 1 with slit 2 closed, and through slit 2 with slit 1 closed, respectively. We assume, with von Neumann, that the detector also obeys quantum physics, with $|ready>$ denoting the "ready" state of the detectors, and $|1>$ and $|2>$ denoting the "clicked" states of each detector. Then the evolution of the composite quantum + detector system, when the quantum passes through slit *i* alone (with the other slit closed), is of the form $|\psi_i> |ready> \rightarrow |\psi_i> |i>$ $(i=1,2)$ (assuming, with von Neumann, that these are "ideal" processes that don't disturb the state of the quantum).

With both slits open, the single quantum approaching the slits is described by a superposition that's extended over both slits:

$$(|\psi_1> + |\psi_2>)/\surd 2 \equiv |\psi>. \qquad (4)$$

Linearity of the time evolution implies that the composite system's evolution during detection at the slits is

$$|\psi> |ready> \rightarrow (|\psi_1> |1> + |\psi_2> |2>)/\surd 2 \equiv |\Psi_{slits}>. \qquad (5)$$

The "measurement state" $|\Psi_{slits}>$ involves both spatially distinguishable detector states $|j>$. It is a "Bell state" of nonlocal entanglement between the quantum and the detector (Ref. 57, pp. 29, 32). If the detectors are reliable, there must be zero probability of finding detector $i$ in the state $|i>$ when detector $j \neq i$ is in its clicked state $|j>$, so $|1>$ and $|2>$ are orthogonal and we assume they are normalized.

It's mathematically convenient to form the pure state density operator

$$\rho_{slits} \equiv |\Psi_{slits}><\Psi_{slits}|, \tag{6}$$

and to form the reduced density operator for the quantum alone by tracing over the detector:

$$\rho_{q\ slits} = Tr_{detector}(\rho_{slits}) = (|\psi_1><\psi_1| + |\psi_2><\psi_2|)/2\ . \tag{7}$$

Eq. (7) has a simple interpretation: Even though the quantum is in the entangled superposition Eq. (5), the result of any experiment involving the quantum *alone* will come out precisely as though the quantum were in one of the pure states $|\psi_1>$ or $|\psi_2>$ with probabilities of 1/2 for each state (Ref. 57). In particular, Eq. (7) predicts that the quantum does not interfere with itself, i.e. there are no interferences between $|\psi_1>$ and $|\psi_2>$. This of course agrees with observation: When detectors provide "which path" information, the interference pattern (i.e. the evidence that the quantum came through both slits) vanishes. The quantum is said to "decohere" (Ref. 57).

To clearly see the field nature of the measurement, suppose there is a "which slit" detector only at slit 1 with no detector at slit 2. Then $|\psi_i> |ready> \rightarrow |\psi_i> |i>$ holds only for $i=1$, while for $i=2$ we have $|\psi_2> |ready> \rightarrow |\psi_2> |ready>$. The previous analysis still holds, provided the "clicked" state $|1>$ is orthogonal to the unclicked state $|ready>$ (i.e. if the two states are distinguishable with probability 1). The superposition Eq. (4) evolves just as before, and Eq. (7) still describes the quantum alone just after measurement. So the experiment is unchanged by removal of one slit detector. Even though there is no detector at slit 2, when the quantum comes through slit 2 it still encodes the presence of a detector at slit 1. This is nonlocal, and it tells us that the quantum extends over both slits, i.e. the quantum is a field, not a particle.

Thus the experiment (Sec. IV A) and the theory both imply each quantum comes through both slits when both slits are open with no detectors, but through one slit when there is a detector at either slit, just as we expect a field (but not a particle) to do.

## C. Theory, at the detecting screen

We'll see that the above analysis at the slits carries over at the detecting screen, with the screen acting as detector.

The screen is an array of small but macroscopic detectors such as single photographic grains. Suppose one quantum described by Eq. (4) passes through the slits and approaches the screen. Expanding in position eigenstates, just before interacting with the screen the quantum's state is

$$|\psi\rangle = \int |\mathbf{x}\rangle\, d\mathbf{x}\, \langle \mathbf{x}|\psi\rangle = \int |\mathbf{x}\rangle\, \psi(\mathbf{x})\, d\mathbf{x} \qquad (8)$$

where the integral is over the 2-dimensional screen, and $\psi(\mathbf{x})$ is the Schroedinger field. Eq. (8) is a (continuous) superposition over position eigenstates, just as Eq. (4) is a (discrete) superposition over slit eigenstates. Both superpositions are extended fields.

Rewriting Eq. (8) in a form that displays the quantum's superposition over the non-overlapping detection regions,

$$|\psi\rangle = \Sigma_i \int_i |\mathbf{x}\rangle\, \psi(\mathbf{x})\, d\mathbf{x} \equiv \Sigma_i A_i\, |\psi_i\rangle \qquad (9)$$

where $|\psi_i\rangle \equiv (1/A_i)\int_i |\mathbf{x}\rangle\, \psi(\mathbf{x})\, d\mathbf{x}$ and $A_i \equiv [\int_i |\psi(\mathbf{x})|^2\, d\mathbf{x}]^{1/2}$. The detection regions are labeled by "i" and the $|\psi_i\rangle$ form an orthonormal set. Eq. (9) is analogous to Eq. (4).

The detection process at the screen is represented by the analog of Eq. (5):

$$|\psi\rangle\, |ready\rangle \rightarrow \Sigma_i\, A_i\, |\psi_i\rangle\, |i\rangle \equiv |\Psi_{screen}\rangle \qquad (10)$$

where $|i\rangle$ represents the "clicked" state of the $i$th detecting region whose output can be either "detection" or "no detection" of the quantum. Localization occurs at the time of this click. Each region i responds by interacting or not interacting, with just one region registering an interaction because a quantum must give up all, or none, of it's energy. As we'll see in Sec. VI C, these other sections of the screen actually register the vacuum--a physical state that can entangle nonlocally with the registered quantum. The non-locality inherent in the entangled superposition state $|\Psi_{screen}\rangle$ has been verified by Bell-type measurements (Sec. VI C). As was the case for detection at the slits (Eq. (5)), Eq. (10) represents the mechanism by which the macro world registers the quantum's impact on the screen.

The argument from Eq. (10) goes through precisely like the argument from Eq. (5) to Eq. (7). The result is that, assuming the states |i> are reliable detectors, the reduced density operator for the quantum alone is

$$\rho_{q\ screen} = \Sigma_i\ |\psi_i\rangle A_i^2 \langle\psi_i|. \tag{11}$$

Eq. (11) tells us that the quantum is registered *either* in region 1 *or* region 2 *or* .... It's this "all or nothing" nature of quantum interactions, rather than any presumed particle nature of quanta, that produces the particle-like interaction regions in Figures 1 and 2.

In summary, "only spatial fields must be postulated to form the fundamental objects to be quantized, ...while apparent 'particles' are a mere consequence of decoherence" [i.e. of localization by the detection process]. [58]

## V. RELATIVISTIC QUANTUM PHYSICS

NRQP (Sec. IV) is not the best basis for analyzing field-particle duality. The spontaneous energy fluctuations of quantum physics, plus SR's principle of mass-energy equivalence, imply that quanta, be they fields or particles, can be created or destroyed. Since relativistic quantum physics was invented largely to deal with such creation and destruction, one might expect relativistic quantum physics to offer the deepest insights into fields and particles.

Quantum physics doesn't fit easily into a special-relativistic framework. As one example, we saw in Sec. III A that photons (relativistic phenomena for sure) cannot be quantum point particles because they don't have position eigenstates.

A more striking example is nonlocality, a phenomenon shown by Einstein, Podolsky, and Rosen,[59] and more quantitatively by John Bell,[60] to inhere in the quantum foundations. Using Bell's inequality, Aspect, Clauser and others showed rather convincingly that nature herself is nonlocal and that this would be true even if quantum physics were not true. [61] The implication is that, by altering the way she measures one of the quanta in an experiment involving two entangled quanta, Alice in New York City can instantly (i.e. in a time too short to allow for signal propagation) change the outcomes observed when Bob measures the other quantum in Paris. This sounds like it violates the special-relativistic prohibition on super-luminal signaling, but quantum physics manages to avoid a contradiction by camouflaging the signal so that Alice's measurement choice is "averaged out" in the statistics of Bob's observations in such a way that Bob detects no change in the statistics of his experiment.[62] Thus Bob receives no signal, even though nonlocality changes his observed results. Quantum physics' particular mixture of uncertainty and non-locality preserves consistency with SR. It's only when Alice and Bob later compare their data that they can spot correlations showing that Alice's change of measurement procedure altered Bob's outcomes. Quantum physics must thread a fine needle, being "weakly local" in order to prevent superluminal signaling but, in order to allow quantum non-locality, not "strongly local" (Ref. 62). Quantum field spreading can transmit information and is limited by the speed of light, while non-local effects are related to superluminal field collapse and cannot transmit information lest they violate SR.

When generalizing NRQP to include such relativistic quantum phenomena as creation and destruction, conflicts with SR can arise unless one generalizes carefully. Hegerfeldt[63] and Malament[64] have each presented rigorous "no-go theorems" demonstrating that, if one assumes a universe containing particles, then the requirements of SR and quantum physics lead to contradictions. This supports the "widespread (within the physics community) belief that the only relativistic quantum theory is a theory of fields."[65] Neither theorem assumes QFT. They assume only SR and the general principles of quantum physics, plus broadly inclusive definitions of what one means by a "particle." Each then derives a contradiction, showing that there can be no particles in any theory obeying both SR and quantum physics. I will present only Hegerfeldt's theorem here, because it is the more intuitive of the two, and because Malement's theorem is more subject to difficulties of interpretation.

Hegerfeldt shows that any free (i.e. not constrained by boundary conditions or forces to remain for all time within some finite region) relativistic quantum "particle" must, if it's localized to a finite region to begin with, *instantly* have a positive probability of being found an arbitrarily large distance away. But this turns out to violate Einstein causality (no superluminal signaling). The conclusion is then that an individual free quantum can never--not even for a single instant--be localized to any finite region.

More specifically, a presumed particle is said to be "localized" at $t_o$ if it is prepared in such a way as to ensure that it will upon measurement be found, with probability 1, to be within some arbitrarily large but finite region $V_o$ at $t_o$. Hegerfeldt then assumes two conditions: First, the presumed particle has quantum states that can be represented in a Hilbert space with unitary time-development operator $U_t = exp(-iHt)$, where $H$ is the energy operator. Second, the particle's energy spectrum has a lower bound. The first condition says that the particle obeys standard quantum dynamics. The second says that the Hamiltonian that drives the dynamics cannot provide infinite energy by itself dropping to lower and lower energies. Hegerfeldt then proves that a particle that is localized at $t_o$ is not localized at any $t > t_o$. See Ref. 63 for the proof. It's remarkable that even localizability in an arbitrarily large finite region can be so difficult for a relativistic quantum particle--its probability amplitude spreads instantly to infinity.

Now here is the contradiction: Consider a particle that is localized within $V_o$ at $t_0$. At any $t > t_0$, there is then a nonzero probability that it will be found at any arbitrarily large distance away from $V_o$. This is not a problem for a non-relativistic theory, and in fact such instantaneous spreading of wavefunctions is easy to show in NRQP.[66] But in a relativistic theory, such instantaneous spreading contradicts relativity's prohibition on superluminal transport and communication, because it implies that a particle localized on Earth at $t_0$ could, with nonzero probability, be found on the moon an arbitrarily short time later. We conclude that "particles" cannot ever be localized. To call a thing a "particle" when it cannot ever be localized is surely a gross misuse of that word.

Because QFTs reject the notion of position observables in favor of parameterized field observables (Sec. III), QFTs have no problem with Hegerfeldt's theorem. In QFT *interactions*, including creation and destruction, occur at specific locations $\boldsymbol{x}$, but the fundamental objects of the theory, namely the fields, do not have positions because they are infinitely extended.

Summarizing: even under a broadly inclusive definition of "particle," quantum particles conflict with Einstein causality.

## VI. THE QUANTUM VACUUM

The Standard Model, a QFT, is today the favored way of looking at relativistic quantum phenomena. In fact, QFT is "the only known version of relativistic quantum theory." [67] Since NRQP can also be expressed as a QFT,[68] all of quantum physics can be expressed consistently as QFTs. We've seen (Sec. V) that quantum particles conflict with SR. This suggests (but doesn't prove) that QFTs are the only logically consistent version of relativistic quantum physics. [69] Thus it appears that QFTs are the natural language of quantum physics.

Because it has energy and non-vanishing expectation values, the QFT vacuum is embarrassing for particle interpretations. If one believes particles to be the basic reality, then what is it that has this energy and these values in the state that has no particles? [70] Because it is the source of empirically verified phenomena such as the Lamb shift, the Casimir effect, and the electron's anomalous magnetic moment, this "state that has no particles" is hard to ignore. This Section discusses QFT vacuum phenomena that are difficult to reconcile with particles. Sec. VI A discusses the quantum vacuum itself. The remaining parts are implications of the quantum vacuum. The Unruh effect (Sec. VI B), related to Hawking radiation, has not yet been observed, while single-quantum nonlocality (Sec. VI C) is experimentally confirmed.

On the other hand, we do not yet really understand the quantum vacuum. The most telling demonstration of this is that the most plausible theoretical QFT estimate of the energy density of the vacuum implies a value of the cosmological constant that is some 120 orders of magnitude larger than the upper bound placed on this parameter by astronomical observations. Possible solutions, such as the anthropic principle, have been suggested, but these remain speculative.[71]

## A. The necessity for the quantum vacuum[72]

Both theory and experiment demonstrate that the quantized EM field can never be sharply (with probability one) zero, but rather that there must exist, at every spatial point, at least a randomly fluctuating "vacuum field" having no quanta. Concerning the theory, recall (Sec. III) that a quantized field is equivalent to a set of oscillators. An actual mechanical oscillator cannot be at rest in its ground state because this would violate the uncertainty principle; its ground state energy is instead $hf/2$. Likewise, each field oscillator must have a ground state where it has energy but no excitations. In the "vacuum state," where the number of excitations $N_{\mathbf{k}}$ is zero for every mode $\boldsymbol{k}$, the expectation values of $\boldsymbol{E}$ and $\boldsymbol{B}$ are zero yet the expectation values of $E^2$ and $B^2$ are not zero. Thus the vacuum energy arises from random "vacuum fluctuations" of $\boldsymbol{E}$ and $\boldsymbol{B}$ around zero.

As a second more direct argument for the necessity of EM vacuum energy, consider a charge $e$ of mass $m$ bound by an elastic restoring force to a large mass of opposite charge. The equation of motion for the Heisenberg-picture position operator $\boldsymbol{x}(t)$ has the same form as the corresponding classical equation, namely

$$d^2\boldsymbol{x}/dt^2 + \omega_o^2\boldsymbol{x} = (e/m)[\boldsymbol{E}_{rr}(t) + \boldsymbol{E}_o(t)]. \qquad (12)$$

Here, $\omega_o$ is the oscillator's natural frequency, $\boldsymbol{E}_{rr}(t)$ is the "radiation reaction" field produced by the charged oscillator itself, $\boldsymbol{E}_o(t)$ is the external field, and it's assumed that the spatial dependence of $\boldsymbol{E}_o(t)$ can be neglected. It can be shown that the radiation reaction has the same form as the classical radiation reaction field for an accelerating charged particle, $\boldsymbol{E}_{rr}(t) = (2e/3c^3)\, d^3\boldsymbol{x}/dt^3$, so Eq. (12) becomes

$$d^2\boldsymbol{x}/dt^2 + \omega_o^2\boldsymbol{x} - (2e^2/3mc^3)d^3\boldsymbol{x}/dt^3 = (e/m)\boldsymbol{E}_o(t). \qquad (13)$$

If the term $\boldsymbol{E}_o(t)$ were absent, Eq. (13) would become a dissipative equation with $\boldsymbol{x}(t)$ exponentially damped, and commutators like $[z(t), p_z(t)]$ would approach zero for large t, in contradiction with the uncertainty principle and in contradiction with the unitary time development of quantum physics according to which commutators like $[z(t), p_z(t)]$ are time-independent. Thus $\boldsymbol{E}_o(t)$ cannot be absent for quantum systems. Furthermore, if $\boldsymbol{E}_o(t)$ is the vacuum field then commutators like $[z(t), p_z(t)]$ turn out to be time-independent.

## B. The Unruh effect

QFT predicts that an accelerating observer in vacuum sees quanta that are not there for an inertial observer of the same vacuum. More concretely, consider Mort who moves at constant velocity in Minkowski space-time, and Velma who is uniformly accelerating (i.e. her acceleration is unchanging relative to her instantaneous inertial rest frame). If Mort finds himself in the quantum vacuum, Velma finds herself bathed in quanta--her "particle" detector clicks. Quantitatively, she observes a thermal bath of photons having the Planck radiation spectrum with $kT = ha/4\pi^2c$ where $a$ is her acceleration.[73] This prediction might be testable in high energy hadronic collisions, and for electrons in storage rings.[74] In fact it appears to have been verified years ago in the Sokolov-Ternov effect.[75]

The Unruh effect lies at the intersection of QFT, SR, and general relativity. Combined with the equivalence principle of general relativity, it entails that strong gravitational fields create thermal radiation. This is most pronounced near the event horizon of a black hole, where a stationary (relative to the event horizon) Velma sees a thermal bath of particles that then fall into the black hole, but some of which can, under the right circumstances, escape as Hawking radiation.[76]

The Unruh effect is counterintuitive for a particle ontology, as it seems to show that the particle concept is observer-dependent. If particles form the basic reality, how can they be present for the accelerating Velma but absent for the non-accelerating Mort who observes the same space-time region? But if fields are basic, things fall into place: Both experience the same field, but Velma's acceleration promotes Mort's vacuum fluctuations to the level of thermal fluctuations. The same field is present for both observers, but an accelerated observer views it differently.

## C. Single-quantum nonlocality

Nonlocality is pervasive, arguably *the* characteristic quantum phenomenon. It would be surprising, then, if it were merely an "emergent" property possessed by two or more quanta but not by a single quantum.

During the 1927 Solvay Conference, Einstein noted that "a peculiar action-at-a-distance must be assumed to take place" when the Schroedinger field for a single quantum passes through a single slit, diffracts in a spherical wave, and strikes a detection screen. Theoretically, when the interaction localizes as a small flash on the screen, the field instantly vanishes over the rest of the screen. Supporting de Broglie's theory that supplemented the Schroedinger field with particles, Einstein commented "if one works only with Schroedinger waves, the interpretation of psi, I think, contradicts the postulate of relativity."[77] Since that time, however, the peaceful coexistence of quantum nonlocality and SR has been demonstrated (Refs. 62, 67).

It's striking that Einstein's 1927 remark anticipated single-quantum nonlocality in much the same way that Einstein's EPR paper (Ref. 59) anticipated nonlocality of two entangled quanta. Today, single-quantum nonlocality has a 20-year history that further demonstrates nonlocality as well as the importance of fields in understanding it.

Single-photon nonlocality was first described in detail by Tan *et. al.* in 1991.[78] In this suggested experiment, a single photon passed through a 50-50 beam-splitting mirror (the "source"), with reflected and transmitted beams (the "outputs") going respectively to "Alice" and "Bob." They could be any distance apart and were equipped with beam splitters with phase-sensitive photon detectors attached to these detectors' outputs.

But nonlocality normally involves two entangled quantum entities. With just one photon, what was there to entangle with? If photons are field mode excitations, the answer is natural: the entanglement was between two quantized field modes, with one of the modes happening to be in the vacuum state. Like all fields, each mode fills space, making nonlocality between modes more intuitive than nonlocality between particles: If a space-filling mode were to instantly change states, the process would obviously be non-local. This highlights the importance of thinking of quantum phenomena in terms of fields. [79]

In Tan *et. al.*'s suggested experiment, Alice's and Bob's wave vectors were the two entangled modes. According to QFT, an output "beam" with no photon is an actual physical state, namely the vacuum state $|0>$. Alice's mode having wave vector $\boldsymbol{k}_A$ was then in a superposition $|1>_A+|0>_A$ of having a single excitation and having no excitation, Bob's mode $\boldsymbol{k}_B$ was in an analogous superposition $|1>_B+|0>_B$, and the two superpositions were entangled by the source beam splitter to create a 2-mode composite system in the nonlocal Bell state

$$|\psi> = |1>_A|0>_B + |0>_A|1>_B \qquad (14)$$

(omitting normalization). Note the analogy with Eq. (5): In Eq. (14), Alice and Bob act as detectors for each others' superposed quanta, collapsing (decohering) both quanta. This entangled superposition state emerged from the source; Alice then detected only mode $\boldsymbol{k}_A$ and Bob detected only mode $\boldsymbol{k}_B$. Quantum theory predicted that coincidence experiments would show correlations that violated Bell's inequality, implying nonlocality that cannot be explained classically.

Analogously to Eq. (7), Alice's and Bob's reduced density operators are

$$\left.\begin{aligned} \rho_A &= Tr_B(|\psi\rangle\langle\psi|) = |1\rangle_{A\,A}\langle 1| + |0\rangle_{A\,A}\langle 0| \\ \rho_B &= Tr_A(|\psi\rangle\langle\psi|) = |1\rangle_{B\,B}\langle 1| + |0\rangle_{B\,B}\langle 0|. \end{aligned}\right\} \qquad (15)$$

Each observer has a perfectly random 50-50 chance of receiving 0 or 1, a "signal" containing no information. All coherence and non-locality are contained in the composite state Eq. (14).

This returns us to Einstein's concerns: In the single-photon diffraction experiment (Sec. IV), interaction of the photon with the screen creates a non-local entangled superposition (Eq. (10)) that is analogous (but with N terms) to Eq. (14). As Einstein suspected, this state is odd, nonlocal. Violation of Bell's inequality shows that the analogous state Eq. (14) is, indeed, nonlocal in a way that cannot be interpreted classically.

In 1994, another single-photon experiment was proposed to demonstrate nonlocality without Bell inequalities.[80] The 1991 and 1994 proposals triggered extended debate about whether such experiments really demonstrate nonlocality involving only one photon.[81] The discussion generated three papers describing proposed new experiments to test single-photon nonlocality.[82] One of these proposals was implemented in 2002, when a single-photon Bell state was teleported to demonstrate (by the nonlocal teleportation) the single-photon nonlocality. In this experiment, "The role of the two entangled quantum systems which form the nonlocal channel is played by the EM fields of Alice and Bob. In other words, the *field modes* rather than the photons associated with them should be properly taken as the information and entanglement carriers" (italics in the original).[83] There was also an experimental implementation of a single-photon Bell test based on the 1991 and 1994 proposals.[84]

It was then suggested that the state Eq. (14) can transfer its entanglement to two *atoms* in different locations, both initially in their ground states $|g\rangle$, by using the state Eq. (14) to generate the joint atomic state $|e\rangle_A|g\rangle_B + |g\rangle_A|e\rangle_B$ (note that the vacuum won't excite the atom).[85] Here, $|e\rangle$ represents an excited state of an atom, while $A$ and $B$ now refer to different modes $\boldsymbol{k}_A$ and $\boldsymbol{k}_B$ of a *matter* field (different beam directions for atoms $A$ and $B$). Thus the atoms (i.e. modes $\boldsymbol{k}_A$ and $\boldsymbol{k}_B$) are placed in a nonlocal entangled superposition of being excited and not excited. Since this nonlocal entanglement arises from the single-photon nonlocal state by purely *local* operations, it's clear that the single-photon state must have been nonlocal too. Nevertheless, there was controversy about whether this proposal really represents single-quantum nonlocality.[86]

Another experiment, applicable to photons or atoms, was proposed to remove all doubt as to whether these experiments demonstrated single-quantum nonlocality. The proposal concluded by stating, "This strengthens our belief that the world described by quantum field theory, where fields are fundamental and particles have only a secondary importance, is closer to reality than might be expected from a naive application of quantum mechanical principles." [87]

## VII. CONCLUSION

There are overwhelming grounds to conclude that all the fundamental constituents of quantum physics are fields rather than particles.

Rigorous analysis shows that, even under the broadest definition of "particle," particles are inconsistent with the combined principles of relativity and quantum physics (Sec. V). And photons, in particular, cannot be point particles because relativistic and quantum principles imply that a photon cannot "be found" at a specific location even in principle (Sec. III A). Many relativistic quantum phenomena are paradoxical in terms of particles but natural in terms of fields: the necessity for the quantum vacuum (Sec. VI A), the Unruh effect where an accelerated observer detects quanta while an inertial observer detects none (Sec. VI B), and single-quantum nonlocality where two field modes are put into entangled superpositions of a singly-excited state and a vacuum state (Sec. VI C).

Classical field theory and experiment imply fields are fundamental, and indeed Faraday, Maxwell, and Einstein concluded as much (Sec. II). Merely quantizing these fields doesn't change their field nature. Beginning in 1900, quantum effects implied that Maxwell's field equations needed modification, but the quantized equations were still based on fields (Maxwell's fields, in fact, but quantized), not particles (Sec. III A). On the other hand, Newton's particle equations were replaced by a radically different concept, namely Schroedinger's *field* equation, whose field solution $\Psi(\mathbf{x},t)$ was however inconsistently interpreted as the probability amplitude for finding, upon measurement, a *particle* at the *point* $\mathbf{x}$. The result has been confusion about particles and measurements, including mentally-collapsed wave packets, students going "down the drain into a blind alley," textbooks filled almost exclusively with "particles talk," and pseudoscientific fantasies (Sec. I). The relativistic generalization of Schroedinger's equation, namely Dirac's equation, is clearly a field equation that is quantized to obtain the electron-positron field, in perfect analogy to the way Maxwell's equations are quantized (Sec. III B). It makes no sense, then, to insist that the non-relativistic version of Dirac's equation, namely the Schroedinger equation, be interpreted in terms of particles. After all, the electron-positron field, which fills all space, surely doesn't shrink back to tiny particles when the electrons slow down.

Thus Schroedinger's $\Psi(\mathbf{x},t)$ is a spatially extended field representing the amplitude for an electron (i.e. the electron-positron field) to interact at **x** rather than an amplitude for finding, upon measurement, a particle. In fact, the field $\Psi(\mathbf{x},t)$ *is* the so-called "particle." Fields are all there is.

Analysis of the 2-slit experiment (Sec. IV) shows why, from a particle viewpoint, "nobody knows how it [i.e. the experiment] can be like that": The 2-slit experiment is in fact logically inconsistent with a particle viewpoint. But everything becomes consistent, and students don't get down the drain, if the experiment is viewed in terms of fields.

Textbooks need to reflect that fields, not particles, form our most fundamental description of nature. This can be done easily, not by trying to teach the formalism of QFT in introductory courses, but rather by talking about fields, explaining that there are no particles but only particle-like phenomena caused by field quantization (Ref. 21). In the 2-slit experiment, for example, the quantized field for each electron or photon comes simultaneously through both slits, spreads over the entire interference pattern, and collapses non-locally, upon interacting with the screen, into a small (but still spread out) region of the detecting screen.

Field-particle duality exists only in the sense that quantized fields have certain particle-like appearances: quanta are unified bundles of field that carry energy and momentum and thus "hit like particles;" quanta are discrete and thus countable. But quanta are not particles; they are excitations of spatially unbounded fields. Photons and electrons, along with atoms, molecules, and apples, are ultimately disturbances in a few universal fields.

## ACKNOWLEDGEMENTS

My University of Arkansas colleagues Julio Gea-Banacloche, Daniel Kennefick, Michael Lieber, Surendra Singh, and Reeta Vyas discussed my incessant questions and commented on the manuscript. Rodney Brooks and Peter Milonni read and commented on the manuscript. I also received helpful comments from Stephen Adler, Nathan Argaman, Casey Blood, Edward Gerjuoy, Daniel Greenberger, Nick Herbert, David Mermin, Michael Nauenberg, Roland Omnes, Marc Sher, and Wojciech Zurek. I especially thank the referees for their careful attention and helpful comments.

---

[a] Email ahobson@uark.edu

[1] M. Schlosshauer, *Elegance and Enigma: The Quantum Interviews* (Springer-Verlag, Berlin, 2011).

[2] N. G. van Kampen, "The scandal of quantum mechanics," Am. J. Phys. **76** (11), 989-990 (2008); A. Hobson, "Response to 'The scandal of quantum mechanics,' by N. G. Van Kampen," Am. J. Phys. **77** (4), 293 (2009).

[3] W. Zurek, "Decoherence and the transition from quantum to classical," Phys. Today **44** (10), 36-44 (1991): "Quantum mechanics works exceedingly well in all practical applications. ...Yet well over half a century after its inception, the debate about the relation of quantum mechanics to the familiar physical world continues. How can a theory that can account with precision for everything we can measure still be deemed lacking?"

[4] C. Sagan, *The Demon-Haunted World: Science as a Candle in the Dark* (Random House, New York, 1995), p. 26: "We've arranged a civilization in which most crucial elements profoundly depend on science and technology. We have also arranged things so that almost no one understands science and technology. This is a prescription for disaster. ...Sooner or later this combustible mixture of ignorance and power is going to blow up in our faces."

[5] M. Shermer, "Quantum Quackery," Scientific American **292** (1), 34 (2005).

[6] V. Stenger "Quantum Quackery," Sceptical Inquirer **21** (1), 37-42 (1997). It's striking that this article has, by coincidence, the same title as Ref. 5.

[7] D. Chopra, *Quantum Healing: Exploring the Frontiers of Mind/Body Medicine* (Bantam, New York, 1989).

[8] D. Chopra, *Ageless Body, Timeless Mind: The Quantum Alternative to Growing Old* (Harmony Books, New York, 1993).

[9] B. Rosenblum and F. Kuttner, *Quantum Enigma: Physics Encounters Consciousness* (Oxford University Press, New York, 2006).

[10] S. Weinberg, *Dreams of a Final Theory: The Search for the Fundamental Laws of Nature* (Random House, Inc., New York, 1992): "Furthermore, all these particles are bundles of the energy, or quanta, of various sorts of fields. A field like an electric or magnetic field is a sort of stress in space... The equations of a field theory like the Standard Model deal not with particles but with fields; the particles appear as manifestations of those fields" (p. 25).

[11] S. Weinberg, *Facing Up: Science and its Cultural Adversaries* (Harvard University Press, Cambridge, MA, 2001): "Just as there is an electromagnetic field whose energy and momentum come in tiny bundles called photons, so there is an electron field whose energy and momentum and electric charge are found in the bundles we call electrons, and likewise for every species of elementary particles. The basic ingredients of nature are fields; particles are derivative phenomena."

[12] R. Mills, *Space, Time, and Quanta: An Introduction to Modern Physics* (W. H. Freeman, New York, 1994), Chp 16: "The only way to have a consistent relativistic theory is to treat *all* the particles of nature as the quanta of fields, like photons. Electrons and positrons are to be treated as the quanta of the electron-positron field, whose 'classical' field equation, the analog of Maxwell's equations for the EM field, turns out to be the Dirac equation, which started life as a relativistic version of the single-particle

Schroedinger equation. …This approach now gives a unified picture, known as *quantum field theory*, of all of nature."
[13] F. Wilczek, "Mass Without Mass I: Most of Matter," Physics Today **52** (11), 11-13 (1999): "In quantum field theory, the primary elements of reality are not individual particles, but underlying fields. Thus, e.g., all electrons are but excitations of an underlying field, ..the electron field, which fills all space and time.
[14] M. Redhead, "More ado about nothing," Foundations of Physics **25** (1), 123-137 (1995): "Particle states are never observable--they are an idealization which leads to a plethora of misunderstandings about what is going on in quantum field theory. The theory is about fields and their local excitations. That is all there is to it."
[15]A. Zee, *Quantum Field Theory in a Nutshell* (Princeton University Press, Princeton, NJ, 2003), p 24: "We thus interpret the physics contained in our simple field theory as follows: In region 1 in spacetime there exists a source that sends out a 'disturbance in the field,' which is later absorbed by a sink in region 2 in spacetime. Experimentalists choose to call this disturbance in the field a particle of mass *m*."
[16] F. Wilczek, "The persistence of ether," Physics Today **52** (1), 11-13 (1999).
[17] F. Wilczek, "Mass Without Mass II: The Medium Is the Massage," Physics Today **53** (1), 13-14 (2000).
[18] F. Wilczek, *The Lightness of Being: Mass, Ether, and the Unification of Forces* (Basic Books, New York, 2008).
[19] R. Brooks, *Fields of Color: The Theory That Escaped Einstein* (Rodney A. Brooks, Prescott, AZ, 2nd ed. 2011). This is a lively history of classical and quantum fields, with many quotations from leading physicists, organized to teach quantum field theory to the general public.
[20] P. R. Wallace, *Paradox Lost: Images of the Quantum* (Springer-Verlag, New York, 1996).
[21] A. Hobson, "Electrons As Field Quanta: A Better Way to Teach Quantum Physics in Introductory General Physics Courses," Am. J. Phys. **73**, 630-634 (2005); "Teaching quantum physics without paradoxes," The Physics Teacher **45**, 96-99 (Feb. 2007); "Teaching quantum uncertainty," The Physics Teacher **49**, 434-437 (2011); "Teaching quantum nonlocality," The Physics Teacher **50**, 270-273 (2012).
[22] A. Hobson, *Physics: Concepts & Connections* (Addison-Wesley/Pearson, San Francisco, 2010).
[23] R. Feynman, *The Character of Physical Law* (The MIT Press, Cambridge, MA, 1965), p. 129. Feynman also says, in the same lecture, "I think I can safely say that nobody understands quantum mechanics."
[24] A. Janiak, *Newton: Philosophical Writings* (Cambridge University Press, Cambridge, 2004), p. 102.
[25] N. J. Nersessian, *Faraday to Einstein: Constructing Meaning in Scientific Theories,* (Martinus Nijhoff Publishers, Boston, 1984), p. 37. The remainder of Sec. III relies strongly on this book.
[26] S. Weinberg, *Facing Up: Science and its Cultural Adversaries* (Harvard University Press, Cambridge, MA, 2001), p. 167: "Fields are conditions of space itself, considered apart from any matter that may be in it."

[27] This argument was Maxwell's and Einstein's justification for the reality of the EM field. R. H. Stuewer, Ed., *Historical and Philosophical Perspectives of Science,* (Gordon and Breach, New York, 1989), p. 299.
[28] A. Einstein, "Maxwell's influence on the development of the conception of physical reality," in *James Clerk Maxwell: A Commemorative Volume 1831-1931* (The Macmillan Company, New York, 1931), pp. 66-73.
[29] A. Einstein, "Zur Elektrodynamik bewegter Koerper," Annalen der Physik **17**, 891-921 (1905).
[30] I. Newton, *Optiks* (4th edition, 1730): "It seems probable to me that God in the beginning formed matter in solid, massy, hard, impenetrable, movable particles ...and that these primitive particles being solids are incomparably harder than any porous bodies compounded of them, even so hard as never to wear or break in pieces...."
[31] R. Brooks, author of Ref. 19, private communication.
[32] J. A. Wheeler and R. P. Feynman, "Interaction With the Absorber as the Mechanism of Radiation," Revs. Mod. Phys. **17**, 157-181 (1945).
[33] P. A. M. Dirac, "The quantum theory of the emission and absorption of radiation," Proceedings of the Royal Society **A114**, 243-267 (1927).
[34] M. Kuhlmann, *The Ultimate Constituents Of The Material World: In Search Of An Ontology For Fundamental Physics* (Ontos Verlag, Heusenstamm, Germany, 2010), Chp. 4; a brief but detailed history of QFT.
[35] The first comprehensive account of a general theory of quantum fields, in particular the method of canonical quantization, was presented in W. Heisenberg and W. Pauli, "Zur quantendynamik der Wellenfelder, *Zeitschrift fuer Physik* **56**, 1-61 (1929).
[36] E. G. Harris, *A Pedestrian Approach to Quantum Field Theory* (Wiley-Interscience, New York, 1972).
[37] More precisely, there are two vector modes for each non-zero $\boldsymbol{k}$, one for each possible field polarization direction, both perpendicular to $\boldsymbol{k}$. See Ref. 36 for other details.
[38] For example, L. H. Ryder, *Quantum Field Theory* (Cambridge University Press, Cambridge, 1996), p 131: "This completes the justification for interpreting $N(k)$ as the number operator and hence for the particle interpretation of the quantized theory."
[39] H. D. Zeh, "There are no quantum jumps, nor are there particles!" Phys. Lett. A **172**, 189-195 (1993): "All particle aspects observed in measurements of quantum fields (like spots on a plate, tracks in a bubble chamber, or clicks of a counter) can be understood by taking into account this decoherence of the relevant local (i.e. subsystem) density matrix."
[40] C. Blood, "No evidence for particles," http://arxiv.org/pdf/0807.3930.pdf: "There are a number of experiments and observations that appear to argue for the existence of particles, including the photoelectric and Compton effects, exposure of only one film grain by a spread-out photon wave function, and particle-like trajectories in bubble chambers. It can be shown, however, that all the particle-like phenomena can be explained by using properties of the wave functions/state vectors alone. Thus there is no evidence for particles. Wave-particle duality arises because the wave functions alone have both wave-like and particle-like properties."
[41] T.D. Newton and E.P. Wigner, "Localized states for elementary systems," Revs. Mod. Phys. **21** (3), 400-406 (1949).

[42] I. Bialynicki-Birula and Z. Bialynicki-Birula, "Why photons cannot be sharply localized," Phys. Rev. A **79**, 032112 (2009).
[43] L. Mandel and E. Wolf, *Optical Coherence and Quantum Optics* (New York, Cambridge UP, 1995).
[44] R. E. Peierls, *Surprises in Theoretical Physics* (Princeton UP, 1979), pp 12-14.
[45] M.G. Raymer and B.J. Smith, "The Maxwell wave function of the photon," SPIE Conference on Optics and Photonics, San Diego, Aug 2005, Conf #5866: The Nature of Light.
[46] Molecules, atoms, and protons are "composite fields" made of the presumably fundamental standard model fields.
[47] The Dirac field is a 4-component relativistic "spinor" field $\Psi_i(\mathbf{x}, t)$ $(i = 1, 2, 3, 4)$.
[48] R. P. Feynman, R. B. Leighton, M. Sands, *The Feynman Lectures on Physics,* Vol. I (Addison-Wesley Publishing Co. Reading, MA, 1963), p. 37-2: "[The 2-slit experiment is] a phenomenon which is impossible, *absolutely* impossible, to explain in any classical way, and which has in it the heart of quantum mechanics. In reality, it contains the *only* mystery. We cannot explain the mystery in the sense of 'explaining' how it works. We will *tell* you how it works. In telling you how it works we will have told you about the basic peculiarities of all quantum mechanics." The italics are in the original.
[49] Nick Herbert, *Quantum Reality: Beyond the new physics* (Doubleday, New York, 1985), pp. 60-67; conceptual discussion of the wave-particle duality of electrons.
[50] Wolfgang Rueckner and Paul Titcomb, "A lecture demonstration of single photon interference," Am J. Phys. **64** (2), 184-188 (1996). Images courtesy of Wolfgang Rueckner, Harvard University Science Center.
[51] A. Tonomura, J. Endo, T. Matsuda, T. Kawasaki, and H. Exawa, "Demonstration of single-electron buildup of an interference pattern," Am. J. Phys. **57** (2), 117-120 (1989).
[52] Michler et al, "A quantum dot single-photon turnstile device," Science **290**, 2282-2285.
[53] For a more formal argument, see A. J. Leggett, "Testing the limits of quantum mechanics: motivation, state of play, prospects," J. Phys: Condensed Matter **14**, R415-R451 (2002).
[54] P. A. M. Dirac, *The Principles of Quantum Mechanics* (Oxford at the Clarendon Press, Oxford, 3rd edition 1947), p. 9. The quoted statement appears in the 2nd, 3rd, and 4th editions, published respectively in 1935, 1947, and 1958.
[55] J. von Neumann, *The Mathematical Foundations of Quantum Mechanics* (Princeton University Press, Princeton, 1955), p. 351.
[56] The central feature of this analysis, namely how decoherence localizes the quantum, was first discussed in W. K. Wootters and W. H. Zurek, "Complementarity in the double-slit experiment: Quantum nonseparability and a quantitative statement of Bohr's principle," Phys. Rev. D **19**, 473-484 (1979).
[57] M. Schlosshauer, *Decoherence and the Quantum-to-Classical Transition* (Springer Verlag, Berlin, 2007), pp. 63-65.
[58] H. D. Zeh, "There is no 'first' quantization," Phys. Lett. A **309**, 329-334 (2003).
[59] A. Einstein, B. Podolsky, and N. Rosen, "Can quantum-mechanical description of physical reality be considered complete?", Phys. Rev. **47** (10), 777-780 (1935).
[60] J. Bell, "On the Einstein Podolsky Rosen Paradox," Physics **1** (3), 195-200 (1964).

[61] A. Aspect, "To be or not to be local," Nature **446**, 866-867 (2007); S. Groblacher, A. Zeilinger, et al, "An experimental test of nonlocal realism," Nature **446**, 871-875 (2007); G. C. Ghirardi, "The interpretation of quantum mechanics: where do we stand?" Fourth International Workshop DICE 2008, Journal of Physics: Conf Series **174** 012013 (2009); T. Norsen, "Against 'realism'," Found. Phys. **37**, 311-340 (2007); D. V. Tansk, "A criticism of the article, 'An experimental test of nonlocal realism'," arXiv 0809.4000 (2008);

[62] L. E. Ballentine and J. P. Jarrett, "Bell's theorem: Does quantum mechanics contradict relativity?" Am. J. Phys. **55**, 696-701 (Aug 1987).

[63] Gerhard C. Hegerfeldt, "Particle localization and the notion of Einstein causality," in *Extensions of Quantum Theory 3,* edited by A. Horzela and E. Kapuscik (Apeiron, Montreal, 2001), pp. 9-16; "Instantaneous spreading and Einstein causality in quantum theory," Annalen der Physik **7**, 716-725 (1998); "Remark on causality and particle localization," PR D **10** (1974), 3320-3321.

[64] D. B. Malament, "In defense of dogma: why there cannot be a relativistic QM of localizable particles." *Perspectives on quantum reality* (Kluwer Academic Publishers, 1996, Netherlands), pp. 1-10. See also Refs. 34 and 65.

[65] H. Halvorson and R. Clifton, "No place for particles in relativistic quantum theories?" Philosophy of Science **69**, 1-28 (2002).

[66] Rafael de la Madrid, "Localization of non-relativistic particles," International Journal of Theoretical Physics, **46**, 1986-1997 (2007). Hegerfeldt's result for relativistic particles generalizes Madrid's result.

[67] P. H. Eberhard and R. R. Ross, "Quantum field theory cannot provide faster-then-light communication," Found. Phys. Letts. **2**, 127-148 (1989).

[68] In other words, the Schroedinger equation can be quantized, just like the Dirac equation. But the quantized version implies nothing that isn't already in the non-quantized version. See Ref. 36.

[69] S. Weinberg, *Elementary Particles and the Laws of Physics, The 1986 Dirac Memorial Lectures* (Cambridge University Press, Cambridge, 1987), pp. 78-79: "Although it is not a theorem, it is widely believed that it is impossible to reconcile quantum mechanics and relativity, except in the context of a quantum field theory."

[70] Michael Redhead, "A philosopher looks at quantum field theory," in *Philosophical Foundations of Quantum Field Theory*, ed by Harvey R. Brown and Rom Harre (Oxford UP, 1988), pp. 9-23: "What is the nature of the QFT vacuum? In the vacuum state ... there is still plenty going on, as evidenced by the zero-point energy ...[which] reflects vacuum fluctuations in the field amplitude. These produce observable effects ....I am now inclined to say that vacuum fluctuation phenomena show that the particle picture is not adequate to QFT. QFT is best understood in terms of quantized excitations of a field and that is all there is to it."

[71] S. Weinberg, "The cosmological constant problem," Revs. Mod. Phys. **61**, 1-23 (1989).

[72] My main source for Secs. VI A and B is Peter W. Milonni, *The Quantum Vacuum: An Introduction to Quantum Electrodynamics* (Academic Press Limited, London, 1994).

[73] W. G. Unruh, "Notes on black hole evaporation," Phys. Rev. D **14**, 870-892 (1976); P.C.W. Davies. “Scalar production in Schwarzschild and Rindler metrics,” Journal of Physics A **8**, 609 (1975).

[74] S. Barshay and W. Troost, "A possible origin for temperature in strong interactions," Phys. Lett. **73B**, 4 (1978); S. Barshay, H. Braun, J. P. Gerber, and G. Maurer, "Possible evidence for fluctuations in the hadronic temperature," Phys. Rev. D **21**, 7 (1980); J. S. Bell and J. M. Leinaas, "Electrons as accelerated thermometers," Nucl. Phys. **B212**, 131 (1983).
[75] Emil T. Akhmedov and Douglas Singleton, "On the physical meaning of the Unruh effect," arXiv:0705.2525v3 [hep-th], 19 Oct 2007.
[76] B. S. DeWitt, "Quantum gravity: the new synthesis," in S. Hawking and W. Israel, Editors, *General Relativity: An Einstein Centenary* (Cambridge University Press, Cambridge, 1979), pp. 680-745.
[77] M. Jammer *The Philosophy of Quantum Mechanics* (Wiley-Interscience, New York, 1974), pp 115-118.
[78] S. M. Tan, D. F. Walls, and M. J. Collett, "Nonlocality of a single photon," Phys. Rev. Lett. **66,** 252-255 (1991).
[79] M.O. Terra Cunha, J.A. Dunningham, V. Vedral, "Entanglement in single-particle systems," Proc. Royal Soc. A **463**, 2277-2286 (2007): "If we want to increase the breadth of applicability of entanglement, we should think in terms of fields which are a fundamental description of nature. Particles are only a manifestation of certain special configurations of quantum fields. If entanglement is to be considered a fundamental property of nature, and even a resource to be understood and applied, one would like to understand entangled fields."
[80] L Hardy, "Nonlocality of a single photon revisited," Phys. Rev. Lett. **73**, 2277-2286 (1994).
[81] L. Hardy, "N-measurement Bell inequalities, N-atom entangled, states, and the nonlocality of one photon," Phys. Lett. A **160**, 1-8 (1991); E. Santos, "Comment on 'Nonlocality of a single photon'," Phys. Rev. Lett. **68**, 894 (1992); D. M. Greenberger, M. A. Horne, and A. Zeilinger, "Nonlocality of a single photon?" Phys. Rev. Lett. **75**, 2064 (1995); Lev Vaidman, "Nonlocality of a single photon revisited again," Phys. Rev. Lett **75**, 2063 (1995); A. Peres, "Nonlocal effects in Fock space," Phys. Rev. Lett. **74**, 4571 (1995); L. Hardy, "Hardy replies," Phys. Rev. Lett. **75**, 2065-2066 (1995); R. J. C. Spreeuw, "A classical analogy of entanglement," Founds. Phys. **28**, 361-374 (1998);
[82] C. C. Gerry, "Nonlocality of a single photon in cavity QED," Phys. Rev. A **53**, 4583-4586 (1996); H. W. Lee and J. Kim, "Quantum teleportation and Bell's inequality using single-particle entanglement," Phys. Rev. A **63**, 012305 (2000); G. Bjork, P. Jonsson, L. L. Sanchez-Soto, "Single-particle nonlocality and entanglement with the vacuum," Phys. Rev. A **64**, 042106 (2001).
[83] E. Lombardi, F. Sciarrino, S. Popescu, and F. De Martini, "Teleportation of a vacuum--one-photon qubit," Phys. Rev. Lett. **88**, 070402 (2002).
[84] B. Hessmo, P. Usachev, H. Heydari, and G. Bjork, "Experimental demonstration of single photon nonlocality," Phys. Rev. Lett. **92**, 2004.
[85] S. J. van Enk, "Single-particle entanglement," Phys. Rev. A **72**, 064306 (2005).
[86] A. Drezet, "Comment on single-particle entanglement," Phys. Rev. **74**, 026301 (2006).
[87] J. Dunningham and V. Vedral, "Nonlocality of a single particle," Phys. Rev. Lett. **99**, 180404 (2007).